\documentclass[aps,prd,preprintnumbers,superscriptaddress,nofootinbib,twocolumn]{revtex4}%

\usepackage[dvips]{graphicx}
\usepackage{feynmf}
\usepackage{color}
\usepackage{bm,latexsym,amsmath,amssymb,amsfonts,mathrsfs}
\usepackage[utf8]{inputenc}
\usepackage[colorlinks=true,linkcolor=magenta,citecolor=magenta]{hyperref}
\usepackage{ulem}
\newcommand*{\ba}{\begin{eqnarray}}
\newcommand*{\ea}{\end{eqnarray}}

\newcommand*{\mpl}{M_{\rm Pl}}
\newcommand{\simgt}{\lower.5ex\hbox{$\; \buildrel > \over \sim \;$}}
\newcommand{\simlt}{\lower.5ex\hbox{$\; \buildrel < \over \sim \;$}}

\newcommand*{\p}{\partial}

\newcommand*{\gb}{{\bar g}}
\newcommand*{\Lphi}{{\cal L}_\phi}

\newcommand*{\Tphiu}{T^{(\phi)}}
\newcommand*{\Tmu}{T^{({\rm b})}}
\newcommand*{\Ttot}{T^{({\rm m})}}
\newcommand*{\Tdmu}{T^{({\rm c})}}

\newcommand*{\Tdmd}{T_{({\rm c})}}

\def\d{\delta}

\def\({\biggl(}
\def\){\biggr)}
\def\[{\biggl[}
\def\]{\biggr]}

\preprint{RUP-17-19}
\begin{document}

\title{Are redshift-space distortions actually a 
probe of growth of structure?}

\author{Rampei Kimura}
\affiliation{Department of Physics, Tokyo Institute of Technology,
	2-12-1 Ookayama, Meguro-ku, Tokyo 152-8551, Japan}

\author{Teruaki~Suyama}
\affiliation{Research Center for the Early Universe (RESCEU),
Graduate School of Science, The University of Tokyo, Tokyo 113-0033, Japan}

\author{Masahide~Yamaguchi}
\affiliation{Department of Physics, Tokyo Institute of Technology,
	2-12-1 Ookayama, Meguro-ku, Tokyo 152-8551, Japan}

\author{Daisuke~Yamauchi}
\affiliation{Faculty of Engineering, Kanagawa University, Kanagawa-ku,
Yokohama-shi, Kanagawa, 221-8686, Japan}

\author{Shuichiro~Yokoyama}
\affiliation{Department of Physics, Rikkyo University, 3-34-1
	Nishi-Ikebukuro, Toshima, Tokyo 171- 8501, Japan}
	\affiliation{Kavli IPMU (WPI), UTIAS, The University of Tokyo,
Kashiwa, Chiba 277-8583, Japan}

\begin{abstract}
We present an impact of coupling between dark matter
and a scalar field, which might be responsible for dark energy, on measurements of 
redshift-space distortions.
We point out that, 
in the presence of conformal and/or disformal coupling,
linearized continuity and Euler equations for
total matter fluid
significantly deviate from the standard ones
even in the sub-horizon scales.  
In such a case, a peculiar velocity of total matter field
is determined not only by a
logarithmic time derivative of its density perturbation
but also by density perturbations for both dark matter and baryon, 
leading to a large modification of the physical interpretation
of observed data obtained by measurements of redshift-space distortions.  
We reformulate galaxy two-point correlation function in the redshift space
based on the modified continuity and Euler equations. 
We conclude from the resultant formula that the true value of the linear growth rate of
large-scale structure cannot be necessarily constrained by single-redshift measurements
of the redshift-space 
distortions, unless one observes the actual time-evolution of structure.
\end{abstract}

\maketitle

%%%%%%%%%%%%%%%%%%%%%%%%%%%%%%%%%%%%%%%%
{\it Introduction. }
%%%%%%%%%%%%%%%%%%%%%%%%%%%%%%%%%%%%%%%%

The current cosmological observations, such as type Ia supernovae
\cite{Perlmutter:1998hx,Riess:1998cb} and cosmic microwave background
\cite{Ade:2015xua}, indicate the presence of dark matter and dark
energy, which have not been identified yet.
The existence of dark matter is also well established by astrophysical
observations, which indicate dark matter as a non-luminous and
pressure-less fluid with small dispersion velocity \cite{Zwicky1933,Babcock1939,Kahn1959,Clowe:2003tk}.
The dark energy is responsible for explaining the late-time accelerated
expansion of the Universe, and numerous attempts to identify it
have been intensively proposed in many literatures.  One such candidate
is to introduce a scalar degree of freedom as a new contribution to
energy-momentum tensor or modification in a gravitational sector (see
for reviews e.g. \cite{Tsujikawa:2010zza,Clifton:2011jh}).

When the ordinary matter, baryon, directly couples with such a
scalar degree of freedom, 
it induces the fifth force.
While the fifth force between baryonic matter is tightly constrained
by the solar-system experiments \cite{Will:2005va},
this is not true for the dark force
that is active only between dark matter since the solar-system experiments
do not probe such an interaction.
Then, the natural arena for probing such interactions is cosmology.
When additional interaction only between the cold dark matter (CDM) is present, 
the growth rate of the CDM density perturbations would be generically different
from that of the baryon density perturbations.
We then expect that observing the growth of the CDM density perturbation
provides us with rich information about such an interaction.

In the standard treatments, 
the linear growth rate of large-scale structure is mainly obtained by
observing galaxy peculiar velocity field through measurements of
redshift-space distortions (RSDs) in galaxy survey.
On large scales, where the linear perturbation theory is valid, 
the galaxy peculiar velocity field is considered to be identical to the velocity field of the total matter.
Based on the continuity equation, the matter velocities should be given by the logarithmic time derivative of the density field, 
that is, the linear growth rate, $f_{\rm m}(a)$.
Galaxy maps produced by estimating distances from observed radial velocities,
which include components from both the Hubble flow and peculiar velocities driven by the clustering
of matter, show an anisotropic galaxy distribution.
Due to such an effect, the galaxy power spectrum on large scales 
is known to be enhanced by the factor $(1+\beta\mu^2)^2$ (named ``Kaiser formula''), where $\beta\equiv f_{\rm m}/b_{\rm g}$ 
with $b_{\rm g}$ being the linear galaxy bias factor and $\mu$ is the cosine of the
angle between the line of sight and the Fourier momentum~\cite{Kaiserl.1987}.  
Although there is a degeneracy between the growth rate and 
the linear bias factor, this degeneracy can be in principle broken by using e.g., 
higher-order statistics \cite{Scoccimarro:1999ed} and cross-correlations between 
other observables 
\cite{Hashimoto:2015tnv} 
by which the linear bias factor alone can be constrained.
Hence, it is widely believed that measurements of RSDs even at single redshift 
allow direct constraints on the growth rate. 
Moreover, 
several attempts show that the relation between the peculiar velocity and the growth rate for each species,
which is based on
the continuity equation, is valid even for the wide range of 
cosmological scenarios including modified theories of gravity
 (see e.g., \cite{Gleyzes:2015rua}).
However, as we will show below, this relation is not necessarily correct in more general situations.
In this {\it Letter}, we would like to address how the above Kaiser
formula is modified when the CDM couples with the scalar field.  

%%%%%%%%%%%%%%%%%%%%%%%%%%%%%%%%%%%%%%%% 
{\it Setup. } 
%%%%%%%%%%%%%%%%%%%%%%%%%%%%%%%%%%%%%%%%
 
Let us consider the following invertible metric transformation
\cite{Bekenstein:1992pj},
\ba
\overline g_{\mu\nu}= A(\phi, X) g_{\mu\nu} + B(\phi, X)\partial_\mu\phi\partial_\nu \phi\,,
\label{gbar}
\ea
where $g_{\mu\nu}$ is the original frame metric, and $A(\phi, X)$ and
$B(\phi, X)$ are respectively called conformal and disformal factors,
which are functions of the scalar field $\phi$ and its kinetic term $X
\equiv -g^{\mu\nu}\partial_\mu \phi \,\partial_\nu \phi /2$. Here
and hereafter, $\phi$ is a generic scalar field, and we do not specify it
though the case with $\phi$ being responsible for dark energy is the
most interesting. The action is given by
\ba
S
&=&
\int {\rm d}^4 x \sqrt{-g} \left[ {\mpl^2 \over 2}\left( R[g]- 2\Lambda\right) + \Lphi[g,\phi ]\right] +S_{\rm m}\,,
\ea
where $\Lphi$ represents a Lagrangian for scalar field and
$S_{\rm m}$ a total matter action. For simplicity, we consider
the canonical scalar field: $\Lphi=
-\frac{1}{2}(\partial\phi)^2-V(\phi )$ and
assume the scalar field does not modify the
gravitational sector, i.e., the absence of kinetic braiding
\cite{Deffayet:2010qz}. As for the matter sector, we assume
that the baryon is minimally coupled for simplicity while the CDM couples
with the scalar field through the barred metric $\gb_{\mu\nu}$ defined
in (\ref{gbar}).
The total matter action is
thus given by
\ba
S_{\rm m}&=&S_{\rm b} + S_{\rm c}
\notag\\
&=&\int{\rm d}^4x\Bigl[\sqrt{-g}{\cal L}_{\rm b} [g_{\mu\nu}, \psi_{\rm b}] +\sqrt{-\bar g}{\cal L}_{\rm c} [\overline g_{\mu\nu},\psi_{\rm c}]\Bigr]\,,
\ea
where $S_{\rm b}$ and $S_{\rm c}$ represent the actions for baryon
and CDM, respectively.
Due to the non-minimal coupling between the dark matter and the scalar field,
baryonic matter and dark matter do not move in the same way.

The variation with respect to the metric $g^{\mu\nu}$ leads to the
Einstein equations as usual,
\ba
G_{\mu\nu}+\Lambda g_{\mu\nu} = \frac{1}{M_{\rm Pl}^2}\left( \Tmu_{\mu\nu} + \Tdmu_{\mu\nu} + \Tphiu_{\mu\nu}\right) \,.
\label{EOMg}
\ea
Here and hereafter, $T_{\mu\nu}^{(\rm I)} = -\frac{2}{\sqrt{-g}}\frac{\delta S_{\rm
I}}{\delta g^{\mu\nu}}$ and $T_{\mu\nu}^{(\phi)} =
-\frac{2}{\sqrt{-g}}\frac{\delta(\sqrt{-g}{\cal L}_{\phi})}{\delta
g^{\mu\nu}}$. The superscript ${\rm I}$ represents ${\rm b}$,
${\rm c}$ or ${\rm m}$ for baryon, dark matter and total matter,
respectively.
The combination of the energy-momentum tensor for total matter 
$\Ttot_{\mu \nu} := \Tmu_{\mu \nu} + \Tdmu_{\mu \nu}$ and the scalar sector $\Tphiu_{\mu\nu}$ is conserved as
$\nabla^\mu ( \Ttot_{\mu \nu} + \Tphiu_{\mu\nu}) =0\,.$
The energy-momentum conservation for baryon also takes the familiar form, $\nabla^\mu \Tmu_{\mu\nu} =0$.
On the other hand, the energy-momentum tensors for the scalar field and dark matter no longer 
satisfy the conservation law individually, and it rather takes the following form, 
$\nabla^\mu \Tdmu_{\mu\nu} = - \nabla^\mu \Tphiu_{\mu\nu} $.

The scalar equation is given by
\ba
\Box\phi - V_\phi =Q\,,
\label{EOMs}
\ea
where $Q$, which characterizes the coupling between  CDM and the scalar field, is defined as
\ba
Q &\equiv&-\frac{1}{\sqrt{-g}}\frac{\delta (\sqrt{-\bar g}{\cal L}_{\rm c})}{\delta\phi} =\nabla_\mu W^\mu- Z \,,\label{Q}
\ea
with
\ba
Z&=& {1\over 2A} \biggl[
\biggl\{A_\phi +\frac{A_XX(A_\phi -2B_\phi X)}{A-A_XX+2B_XX^2}\biggr\} \Tdmd \nonumber\\
&&+\biggl\{B_\phi +\frac{B_XX(A_\phi -2B_\phi X)}{A-A_XX+2B_XX^2}\biggr\}  \Tdmd^{\mu\nu}\partial_\mu\phi \partial_\nu\phi\label{eq:Z def}
\biggr] ,~~~~~\\
W^\mu&=&{1\over 2A} \biggl[
2B\, \Tdmd^{\mu\nu} \partial_\nu\phi
-\frac{A-2BX}{A-A_XX +2B_XX^2}\nonumber\\
&&~~~~~~~~~\times\bigl(A_X \Tdmd +B_X \Tdmd^{\alpha\beta}\partial_\alpha\phi\partial_\beta\phi\bigr)\partial^\mu\phi\label{eq:W def}
\biggr]\,,
\ea
where $U_\phi =\partial U/\partial\phi$\,, $U_X=\partial U/\partial X$ for $U=A\,,B$\,.
By the use of Eq. (\ref{EOMs}), the energy-momentum conservation for CDM and total matter % (\ref{EMC}) 
can be recast as
\ba
\nabla^\mu\Tdmu_{\mu\nu}=\nabla^\mu \Ttot_{\mu\nu} = - Q\, \partial_\nu\phi \,.
\label{EOMDM}
\ea

%%%%%%%%%%%%%%%%%%%%%%%%%%%%%%%%%%%%%%%%
{\it Basic equations. }
%%%%%%%%%%%%%%%%%%%%%%%%%%%%%%%%%%%%%%%%

We work on a spatially flat FLRW metric in Newtonian gauge,
\ba
{\rm d}s^2=-[1+2\Phi(t, {\bm x})]{\rm d}t^2 + a^2(t) [1-2\Psi(t, {\bm x})]{\rm d}{\bm x}^2\,,
\ea
and define the background and perturbations of the energy-momentum
tensor for the baryon, the dark matter and the total matter as
\ba
&&T_{({\rm I})0 }^{~0} = -\rho_{\rm I}(t)\Bigl[1+\delta_{\rm I} (t,{\bm x})\Bigr]\,, \\
&&T_{({\rm I})i }^{~0} = - \rho_{\rm I}(t) \, \p_i v_{\rm I}(t,{\bm x})\,,
%\\
%&&T^{({\rm I})} {}^i_{~j}=0\,,
\ea
and (otherwise)$=0$
\footnote{Note that the pressureless feature of the
CDM is robust
at least at first order of perturbations even if we take other
definitions of energy momentum tensor such as
$\overline{T}_{\mu\nu}^{({\rm c})}=-(2/\sqrt{-\overline{g}})\delta(\sqrt{-\overline{g}}{\cal L}_{\rm c})/\delta\overline{g}^{\mu\nu}$ and $\widetilde{T}_{\mu\nu}^{({\rm c})}=-(2/\sqrt{-g})\delta(\sqrt{-g}{\cal L}_{\rm c})/\delta g^{\mu\nu}$.}.
Based on these equations, we can find relations as
\ba
\label{eq:totalm}
\delta_{\rm m} &=& \omega_{\rm c} \delta_{\rm c} + \omega_{\rm b} \delta_{\rm b} \,,\\
v_{\rm m} &=& \omega_{\rm c}v_{\rm c} + \omega_{\rm b} v_{\rm b}\,,
\label{eq:totalv}
\ea
where $\omega_{\rm I} = \rho_{\rm I} / \rho_{\rm m}$.
We also split the scalar field as
$\phi(t, {\bm x})\to \phi(t) + \delta\phi(t,{\bm x})$.
The background part of the Einstein equation gives 
\ba
&&H^2 =\frac{1}{3M_{\rm Pl}^2}\Bigl(\rho_{\rm c}+ \rho_{\rm b}+\Lambda +\frac{1}{2}\dot\phi^2+V\Bigr)\,,\\
&&3 H^2+2 \dot{H}  = \frac{1}{M_{\rm Pl}^2}\left( \Lambda -\frac{1}{2}\dot\phi^2 +V\right) \,.
\ea
The background equation of motion for $\phi$ (\ref{EOMs}) yields
\ba
\ddot{\phi }+3 H \dot{\phi}+V_{\phi }=-Q_0\,,
\ea
and the energy-momentum conservation for baryon and CDM lead to the
background equations
\ba
\dot{\rho}_{\rm b}+3 H \rho_{\rm b} &=& 0\,,\\
\label{rhobB}
\dot{\rho}_{\rm c}+3 H \rho_{\rm c} &=&Q_0 \dot{\phi }\,,
\label{rhocB}
\ea
where $Q_0$ is a background value of $Q$. 
We can rewrite $Q_0$ from the
definition \eqref{Q}-\eqref{eq:W def} together with (\ref{rhocB}),
\ba
\frac{\dot\phi}{\rho_{\rm c}}Q_0= \frac{1}{2}\frac{\rm d}{{\rm d}t}\log\Biggl[\frac{(2A-A_X\dot\phi^2+B_X\dot\phi^4)^2}{A-B\dot\phi^2}\Biggr]\,,
\label{Q0}
\ea
where $A$\,, $A_X$\,, $B$ and $B_X$ are evaluated at background fields.

In deriving perturbed equations, we use the quasi-static approximation,
which is applicable when the wavelength of perturbations is well inside
the sound Horizon of the scalar field, $k^{-1} \ll c_s / (a H)$, 
where $c_s$ is the sound speed of the scalar field. 
Although the sound speed of the scalar field generally differs from unity in our setup \cite{KSYYY2}, we assume $c_s = {\cal O}(1)$ for simplicity.
Then we can neglect time-derivative terms of perturbations
while keeping spatial derivative terms~\footnote{ We also neglect the
mass $m_\phi$ of the scalar field, which could be crucial when the mass
of the scalar field is large enough, i.e., $m_\phi \simgt k/a$.  The
full analysis including large mass effects will be reported in the full
paper \cite{KSYYY2}.  }, and we obtain the linearized perturbed Einstein
equations in the Fourier space,
\ba
\frac{k^2}{a^2}  \Psi = \frac{k^2}{a^2} \Phi =  -4\pi G
\rho_{\rm m} \delta_{\rm m} \,.
\ea
The continuity and Euler equations for baryon are the standard form; 
\ba
\dot{\delta}_{\rm b} +{k^2 \over a^2} v_b &=& 0 \,, \\
\dot{v}_b -\Phi &=&0 \,,
\ea
while those for CDM get modified as follows
\ba
\dot\delta_{\rm c}+\frac{k^2}{a^2}v_{\rm c}&=&\frac{\dot\phi}{\rho_{\rm c}}\Bigl(\delta Q-Q_0\delta_{\rm c}\Bigr)\,,\\
\dot v_{\rm c}-\Phi &=&\frac{Q_0}{\rho_{\rm c}}\left(\delta\phi -\dot\phi v_{\rm c}\right)\,,
\ea
where the scalar field perturbation is determined by
\ba
-\frac{k^2}{a^2}\delta\phi =\delta Q\,.
\label{eq:delta phi eq}
\ea
In the quasi-static limit, the most relevant terms in $Q$ can be
extracted as
\ba
	\delta Q
		=Q_0\delta_{\rm c}
			+\left( R_1+R_2\right)\dot\phi\dot\delta_{\rm c}+R_1\dot\phi\frac{k^2}{a^2}v_{\rm c}+R_2\frac{k^2}{a^2}\delta\phi
	\,,\label{eq:delta Q}
\ea
where
\ba
	&&R_1=\frac{B\rho_{\rm c}}{A}\,,\\
	&&R_2=-\frac{(A-B\dot\phi^2)(A_X-B_X\dot\phi^2)\rho_{\rm c}}{A(2A-A_X\dot\phi^2 +B_X\dot\phi^4)}
	\,.
\ea
Physical meanings of these functions are as follows: 
$R_1$
characterizes the strength of the modification of the CDM 
velocity field from the disformal coupling, and 
$R_2$ represents the contribution from the scalar field
perturbation in $\delta Q$.  
We emphasize that the $R_2$ term gives the non-vanishing contributions only if the conformal and/or
disformal factors depend on the kinetic term. 

After eliminating the scalar field perturbations by using \eqref{eq:delta phi eq}, we can rewrite the modified continuity 
equation in terms of the CDM density contrast and %peculiar
velocity field as
\ba
	\left( 1-\Upsilon_1\right)\left(\dot\delta_{\rm c}+\frac{k^2}{a^2}v_{\rm c}\right)
		=\Upsilon_2\left(\dot\delta_{\rm c}-\frac{Q_0}{\dot\phi}\delta_{\rm c}\right)
	\,,\label{eq:modified continuity eq}
\ea
with
\ba
	\Upsilon_1=\frac{\dot\phi^2}{\rho_{\rm c}}\frac{R_1}{1+R_2}
	\,,\ \ 
	\Upsilon_2=\frac{\dot\phi^2}{\rho_{\rm c}}\frac{R_2}{1+R_2}
	\,.
\ea
In the minimal coupling case ($A=1,B=0$), 
all time dependent coefficients are zero, $Q_0=R_1=R_2=0$. 
When conformal and disformal factors depend only on $\phi$, 
we have $Q_0\neq 0$\,, $R_1\neq 0$, and $R_2=0$.
Thus the continuity equation is the same as the one in the minimal
coupling case.
One can verify this property even for a wider class of scalar-tensor theories~\cite{Gleyzes:2015rua}. 
When at least one of $A$ and $B$ depend on $X$, there arises
a new contribution of $R_2$ in the continuity equation,
and the CDM velocity can significantly differ from the standard case. 
An important implication from these equations is that 
the continuity equation for the total matter fluctuations, \eqref{eq:totalm} and \eqref{eq:totalv}, is given by 
\ba
	\dot\delta_{\rm m}+\frac{k^2}{a^2}v_{\rm m}
                &=& \frac{\dot\phi}{\rho_{\rm m}}\left[
                    \delta Q - Q_0 \left(\omega_{\rm b}\delta_{\rm b} 
                          + \omega_{\rm c}\delta_{\rm c} \right)\right]  \notag\\
		&=& \omega_{\rm c}\frac{\Upsilon_2}{1-\Upsilon_1}\left(\dot\delta_{\rm c}-\frac{Q_0}{\dot\phi}\delta_{\rm c}\right)
	\notag\\
	&&\, 
			+\omega_{\rm b}\frac{Q_0\dot\phi}{\rho_{\rm
			m}}\left(\delta_{\rm c}-\delta_{\rm b}\right)\,,
\label{eq:continuity eq for total matter}
\ea
which differs from the standard form by the presence of the
non-minimal coupling.
We also found that even when the $R_2$ contribution is negligible
the standard form of the continuity equation cannot be reproduced due to the second term
of the right-hand-side in the second equation, which originates from the deviation of the background energy density 
from the standard matter (see Eq.~\eqref{rhocB}).
Therefore, we conclude that there are two possibilities to break the standard relation of the continuity equation for the total matter field:
One comes from the $R_2$ term in the CDM continuity equation, which appears only when the coupling depends on
the kinetic term, 
and the other corresponds to the deviation of the background dynamics from the standard one characterized by $Q_0$.

Combining all the perturbed equations to eliminate velocities as usual, 
we obtain two coupled second-order differential equation for the baryon and CDM density contrasts.
Since the evolution equations for the baryon and CDM density contrasts are independent of the wavenumber $k$,
one can decompose the density contrasts into the (normalized) $k$-independent linear growth factors 
$D_{\rm I}$ 
and initial density contrasts $\delta_0$ for the baryon, CDM and total matter as follows, 
\ba
\d_{\rm I}(t, {\bm k} ) = D_{\rm I} (t) \delta_0 ({\bm k})\,.
\ea
Here, we have chosen the initial time to be much after the time of CMB decoupling ($z \approx 1100$)
but much before the effect of the dark interaction becomes important ($z \sim 1$) and assumed that 
the baryon density contrast has caught up with the CDM density contrast by the initial time.
We also define the growth rate for each species, $f_{\rm I}$, as the logarithmic derivative of the linear growth,
that is $f_{\rm I}(t)\equiv{\rm d}\ln D_{\rm I}/{\rm d}\ln a$.
Rewriting the continuity equations for the baryon, CDM and total matter
in terms of the growth factors, we obtain the suggestive form of velocity potentials
\ba
v_{\rm I}(t,{\bm k})  = - {a^2 H\over k^2} f_{\rm I}^{\rm eff}(t) \delta_{\rm I}(t,{\bm k})\,.
\ea
Although one can easily see $f_{\rm b}^{\rm eff}=f_{\rm b}$,
from Eq.~\eqref{eq:modified continuity eq} the {\it effective} linear growth rate of the CDM, $f_{\rm c}^{\rm eff}$, can significantly differ from the standard one due to the $R_2$ contribution as
\ba
	f_{\rm c}^{\rm eff} 
		= f_{\rm c}-\frac{\Upsilon_2}{1-\Upsilon_1}\left( f_{\rm c}-\frac{Q_0}{H\dot\phi}\right)
		\equiv f_{\rm c}+\Delta f_{\rm c}
	\,.\label{LGR}
\ea
Moreover,  by the use of Eq.~\eqref{eq:continuity eq for total matter}, 
the {\it effective} growth rate of the total matter fluctuations can be written as
\ba
	f_{\rm m}^{\rm eff}
		&=&\frac{\omega_{\rm c}D_{\rm c}f_{\rm c}^{\rm eff}+\omega_{\rm b}D_{\rm b}f_{\rm b}}{\omega_{\rm c}D_{\rm c}+\omega_{\rm b}D_{\rm b}}
	\notag\\
		&=&f_{\rm m}+\omega_{\rm c}\frac{D_{\rm c}}{D_{\rm m}}\Delta f_{\rm c}
			-\omega_{\rm b}\frac{Q_0\dot\phi}{H\rho_{\rm m}}\frac{D_{\rm c}-D_{\rm b}}{D_{\rm m}}
%		\equiv f_{\rm m}+\Delta f_{\rm m}
	\,.\label{eq:f_m^eff}
\ea
with $D_{\rm m}=\omega_{\rm c}D_{\rm c}+\omega_{\rm b}D_{\rm b}$.
Although $f_{\rm m}^{\rm eff}$ is naturally given by the growth-factor-weighted average of the effective growth rates for CDM and baryon, 
it does not in general coincide with $f_{\rm m}$.
As discussed above, its deviation is due to
the non-trivial terms in the CDM continuity equation and the background dynamics 
that produce the second and third terms in \eqref{eq:f_m^eff}.

%%%%%%%%%%%%%%%%%%%%%%%%%%%%%%%%%%%%%%%%
{\it Modified interpretation of Kaiser formula. }
%%%%%%%%%%%%%%%%%%%%%%%%%%%%%%%%%%%%%%%%

In the above investigation, we found that the {\it effective} growth rate $f_{\rm m}^{\rm eff}$  inferred from the peculiar velocities no longer coincides with the actual growth rate $f_{\rm m}$, 
namely measurements of the peculiar velocity field do not necessarily provide the growth rate of clustering directly.
Our example vividly demonstrates that the standard dictionary 
translating the RSDs measurements into the growth rate
is not universal and fails for some classes of theories beyond the $\Lambda$CDM model.
To see the impact of the breaking of the relation between the peculiar velocities and the actual growth rate, 
we now focus on the modification of the Kaiser formula as the simplest and most important observable effect of RSDs.
The generalization to other observables related to the peculiar velocities is straightforward.

In addition to the Hubble expansion, the peculiar velocities of the
galaxies relative to the Hubble expansion distort the distribution of galaxies in the 3-dimensional redshift space, and such effects must carefully be taken into account when
comparing galaxy two-point correlation function with theoretical
predictions \cite{Kaiserl.1987}.  The mapping of the observed redshift
position ${\bm s}$ from the real space position ${\bm x}$ is given by
\ba
{\bm s}={\bm x}+ \frac{v_{\text{g},z}}{aH} \widehat{\bm z}  \,,
\label{eq:s def}
\ea
where $v_{\text{g},z}$ is a line-of-sight component of the peculiar velocity
of a galaxy
and $\widehat{\bm z}$ is a unit vector of
line-of-sight.
In Eq.~\eqref{eq:s def}, we have assumed the plane-parallel approximation,
so that the line-of-sight is taken as a fixed direction, $\widehat{\bm z}$.
Recalling that the number of galaxies in the
infinitesimal volume of both spaces is invariant, the overdensities in
the redshift space $\delta_{ \text{g},s}$ and the real space $\delta_{\text{g}}$
are related through
\ba
\delta_{ \text{g},s} =\delta_{\text{g}}-  \frac{1}{aH}\nabla_z v_{\text{g},z}  \,.
\ea
The galaxy density contrast in the real space, $\delta_{\rm g}$, is related to the total matter density
contrast $\delta_{\rm m}$ 
given by Eq. (\ref{eq:totalm}),
through the standard linear bias model $\delta_{\rm g}=b_{\rm g}\delta_{\rm m}$ on large scales.
The peculiar velocity fields of the galaxies, $v_{\text{g}}$, 
on large scales are expected to be related to
the CDM and baryon fluid velocities, and the explicit relation is determined by
imposing the reasonable physical condition, e.g., momentum conservation law for each galaxy~\cite{Gleyzes:2015rua}.
For simplicity, here, we assume that 
the peculiar velocity fields of galaxies on large scales are given
by the total matter fluid velocities, $v_{\rm m}$, given in Eq. (\ref{eq:totalv}), as in the standard case: 
$v_{\rm g}=v_{\rm m}=-(a^2H/k^2)f_{\rm m}^{\rm eff}\delta_{\rm m}$.

Therefore, the resultant galaxy power spectrum in redshift space is given by 
\ba \label{galps}
P_{\text{g},s} ({\bm k};t)  =\,  {\big( 1+\beta_{\rm eff}(t)\,\mu^2 \big)}^2 P_{\rm g} (k;t) \;,
\ea
where $P_{\rm g}=b_{\rm g}^2\,P_{\rm m}$ is the real-space galaxy
power spectrum, $P_{\rm m}=D_{\rm m}^2P_0$
is the power spectrum for the total matter
density contrast, and
\ba 
	\beta_{\rm eff} &\equiv& \frac{f_{\rm m}^{\rm eff}}{b_{\rm g}}
	\notag\\
	\ \ 
			&=&\beta +\frac{1}{b_{\rm g}D_{\rm m}}\biggl[\omega_{\rm c}D_{\rm c}\Delta f_{\rm c}
			-\omega_{\rm b}\frac{Q_0\dot\phi}{H\rho_{\rm m}}\left( D_{\rm c}-D_{\rm b}\right)\biggr]
	\,.~~~~~\label{eq:beta_eff}
\ea
This is a generalization of the Kaiser formula.  
In fact, in the minimal coupling case, we have $D_{\rm m} = D_{\rm c} = D_{\rm b}$ and $f_{\rm m}^{\rm eff}=f_{\rm m}=f_{\rm c}^{\rm eff} = f_{\rm c} = f_{\rm b}$, 
and hence Eq. (\ref{galps}) is reduced to the standard Kaiser formula. 
However, we found that 
in the presence of the coupling between the CDM and the scalar sector
we have no longer the relation $f_{\rm m}^{\rm eff}=f_{\rm m}$ as we have discussed,
and it means that the RSDs are not trustable probes of growth of structure.
It is notable that the RSDs cannot provide the true value of the growth rate $f_{\rm m}$
even in the simple case where the conformal and disformal factors depend only on $\phi$.
Since in this case, the deviation from the standard formula is
proportional to $Q_0$, this effect is suppressed 
when the background evolution of the dark matter
is almost same as the one of the baryon.
On the other hand, there is a wider room for sizable modification of the standard Kaiser formula
in our general setup;
even when either $Q_0$ or the baryonic contamination is negligibly small, $f_{\rm m}^{\rm eff}$
can differ from $f_{\rm m}$ by ${\cal O}(1)$.
To see this clearly, let us expand the formula (\ref{eq:beta_eff}) in terms of the baryon-CDM ratio 
to neglect the ambiguity from the baryon contribution.
The leading term gives $\beta_{\rm eff}\approx f_{\rm c}^{\rm eff}/b_{\rm g}=f_{\rm c}/b_{\rm g}+\Delta f_{\rm c}/b_{\rm g}$.
This immediately shows 
the single-redshift RSDs measurements can not give a constraint on the linear growth rate $f_{\rm c}$
unless the contributions from the couplings $\Delta f_{\rm c}$ is fixed by using other observables.
This fact demonstrates that one has to keep this new effect in mind when testing 
beyond $\Lambda$CDM theories by the RSDs measurements.
Even if the growth index $\gamma\approx 0.55$ is obtained from RSDs in future galaxy survey, 
it is still possible that the true theory is different from the standard $\Lambda$CDM model.
One way to obtain the actual growth rate of large-scale structure is to directly observe the time-evolution of structure by
e.g. multiple redshift observations of galaxy power spectrum.
In fact, we have a strong degeneracy between the growth of large scale structure and the redshift-dependence of the linear bias. Thus, to measure $f_{\rm m}$ by multiple redshift observations,
we need to fix the bias for each redshift by using other observations, i.e., cross-correlation between the clustering of galaxies and weak lensing (see, e.g., \cite{Hashimoto:2015tnv}).
After evaluating the actual growth rate, one can compare the actual and effective growth rates to constrain the couplings
between the CDM and scalar field.

%%%%%%%%%%%%%%%%%%%%%%%%%%%%%%%%%%%%%%%%
{\it Conclusion. }
%%%%%%%%%%%%%%%%%%%%%%%%%%%%%%%%%%%%%%%%

We have shown that the additional interaction mediated by the scalar field that operates only
between dark matter through conformal and disformal couplings changes the continuity and Euler equations for cosmological perturbations in a non-trivial manner
and investigated its impact on   
RSDs measurements in galaxy survey.
We found that the effects of such
modifications appear even at sub-horizon scales in the presence of
$\phi$ and $X (=-g^{\mu\nu}\partial_\mu \phi \partial_\nu \phi/2)$-dependence of 
the conformal and/or disformal couplings.  
The {\it effective} linear
growth rate, which is inferred from measurements of the peculiar velocities of 
the distributed galaxies, no longer corresponds to the logarithmic time derivative of 
the density perturbation and is rather characterized by both the density
perturbations and their derivatives for each species in general situation.  
In other words, the information of the coupling is encoded in the peculiar velocity fields 
and the true value of the growth rate of large-scale structure cannot 
necessarily be constrained by the single-redshift RSDs measurements.
It can be extracted by using multiple power spectra of the galaxy distribution at different
redshift.  This fact will play a vital role of measuring the linear
growth rate $f_{\rm m}$ by the RSDs measurement, and it will provide us a
rich information of dark matter and dark energy.

%%%%%%%%%%%%%%%%%%%%%%%%%%%%%%%%%%%%%%%%
{\it Acknowlegments}. 
%%%%%%%%%%%%%%%%%%%%%%%%%%%%%%%%%%%%%%%%

We thank Toshifumi Noumi, Masamune Oguri, Jiro Soda, Masahiro Takada, and Kazuhiro Yamamoto for many useful comments and discussions.
This work was supported in part by JSPS Grant-in-Aid for Scientific Research Nos.~JP17K14304 (D.Y.), JP17K14276 (R.K.), JP25287054 (R.K. \& M.Y.), 
JP26610062 (M.Y.), JP15K17632 (T.S.) and JP15K17659 (S.Y.),
MEXT KAKENHI Grant-in-Aid for Scientific Research on Innovative Areas "Cosmic Acceleration"  Nos.~JP15H05888 (M.Y.) and JP16H01103 (S.Y.), 
MEXT KAKENHI Grant Numbers JP17H06359 (T.S.).

%%%%%%%%%%%%%%%%%%%%%%%%%%%%%%%%%%%%%%%%
{\it Note added}. 
%%%%%%%%%%%%%%%%%%%%%%%%%%%%%%%%%%%%%%%%

While this paper was being completed, Ref.~\cite{BW2017} appeared, 
in which the redshift space distortions in the context of interacting dark matter and
vacuum energy are discussed.

\bibliography{references}

\end{document}